\documentclass[conference]{IEEEtran}
\makeatletter
\def\ps@headings{%
\def\@oddhead{\mbox{}\scriptsize\rightmark \hfil \thepage}%
\def\@evenhead{\scriptsize\thepage \hfil \leftmark\mbox{}}%
\def\@oddfoot{}%
\def\@evenfoot{}}
\makeatother
\IEEEoverridecommandlockouts
\usepackage{cite}
\usepackage{amsmath,amssymb,amsfonts}
\usepackage{bm,setspace,mathtools}
\usepackage{algorithm}
\usepackage{algpseudocode}
\usepackage{mathrsfs}
\usepackage{graphicx}
\usepackage{textcomp}
\usepackage{xcolor}
\usepackage{booktabs}
\usepackage{float}
\usepackage{subfigure}
\usepackage{url}  
\usepackage{makecell}    

\newcommand{\system}{\textit{ARGOS}} 

\def\BibTeX{{\rm B\kern-.05em{\sc i\kern-.025em b}\kern-.08em
    T\kern-.1667em\lower.7ex\hbox{E}\kern-.125emX}}
\begin{document}

\title{ARGOS: Anomaly Recognition and Guarding through O-RAN Sensing}

\author{\IEEEauthorblockN{Stavros Dimou}
\IEEEauthorblockA{\textit{Northeastern University} \\
\textit{Boston, MA, USA}\\
dimou.s@northeastern.edu}
\and
\IEEEauthorblockN{Guevara Noubir}
\IEEEauthorblockA{\textit{Northeastern University} \\
\textit{Boston, MA, USA}\\
g.noubir@northeastern.edu}
\vspace{-2.5em}
}
\maketitle

\begin{abstract}

Rogue Base Station (RBS) attacks, particularly those exploiting downgrade vulnerabilities, remain a persistent threat as 5G Standalone (SA) deployments are still limited and User Equipment (UE) manufacturers continue to support legacy network connectivity. 
This work introduces \textit{\system}, a comprehensive O-RAN compliant Intrusion Detection System (IDS) deployed within the Near Real-Time RIC, designed to detect RBS downgrade attacks in real time, an area previously unexplored within the O-RAN context.
The system enhances the 3GPP KPM Service Model to enable richer, UE-level telemetry and features a custom xApp that applies unsupervised Machine Learning models for anomaly detection. 
Distinctively, the updated KPM Service Model operates on cross-layer features extracted from \textit{Modem Layer 1 (ML1)} logs and \textit{Measurement Reports} collected directly from Commercial Off-The-Shelf (COTS) UEs. 
To evaluate system performance under realistic conditions, a dedicated testbed is implemented using Open5GS, srsRAN, and FlexRIC, and validated against an extensive real-world measurement dataset. 
Among the evaluated models, the Variational Autoencoder (VAE) achieves the best balance of detection performance and efficiency, reaching \textit{99.5\% Accuracy} with only \textit{0.6\% False Positives} and minimal system overhead.

\end{abstract}

\begin{IEEEkeywords}
5G, O-RAN, xApps, ML, RBS, IDS
\end{IEEEkeywords}

\section{Introduction}
\label{sec:intro}

The rise of 5G mobile networks represents a major shift in telecommunications, opening the door to innovative applications through faster connectivity and minimal delay.
As the telecom industry nears the midpoint of 5G adoption, a growing number of Mobile Network Operators (MNOs), particularly in Asia and North America, are focusing on network densification and transitioning to 5G Core Standalone (SA) architectures \cite{ookla}. 
This transition is intended to unlock the full potential of 5G SA, with global subscriptions expected to reach 3.6 billion by 2030 \cite{ericsson_mobility}. 
Nonetheless, studies indicate that LTE and earlier mobile networks will continue to be used globally, suggesting that 5G SA will coexist with legacy networks well into the next decade, especially as user devices remain compatible with older cellular technologies.
This raises the problem of backward compatibility with older generations, which continue to expose legacy vulnerabilities.
Downgrade attacks take advantage of this, forcing devices to connect through less secure legacy networks, compromising the connection’s integrity. 
Such attacks are commonly carried out by adversaries using Rogue Base Stations (RBS), which operate as International Mobile Subscriber Identity (IMSI) catchers \cite{neverLetmeDown, catch_me_if, imsi_catch, Imsi_Wild, NewVuln, LookBefore}. 
Thus, as we move beyond 5G and towards 6G, networks must be equipped to detect and classify malicious entities and traffic within their vicinity. 

From both operational and security perspectives, a key advancement in 5G is the adoption of the software-defined Open Radio Access Network (O-RAN) architecture, which introduces a new level of programmability to traditional cellular infrastructures. 
O-RAN redefines the traditionally monolithic and vendor-proprietary RANs by introducing a disaggregated, modular architecture that promotes openness, interoperability, and programmability, as illustrated in Figure \ref{fig:o-ran}. 
Embracing the principles of Software-Defined Networking (SDN), O-RAN enables centralized control and dynamic network optimization across the RAN. 
Central elements of this architecture are the RAN Intelligent Controllers (RICs), located in the Control Plane, both supporting modular, ``plug-and-play" applications, also known as xApps and rApps.
Recent studies have increasingly leveraged the RIC for both network optimization and the implementation of Intrusion Detection Systems (IDS) and Intrusion Prevention Systems (IPS), addressing a range of attack vectors.

However, no prior work has explicitly addressed downgrade attacks by proposing a practical solution aligned with current and future cellular network deployments and smartphone manufacturing constraints.
Moreover, much of the existing research relies on synthetic data or constrained measurements, failing to reflect the characteristics of existing cellular network deployments.
To address these gaps, we introduce \system \footnote{\textit{Argos Panoptes}, the “all-seeing” giant in Greek mythology, had a hundred eyes and symbolized perpetual vigilance—reflecting \system’s continuous monitoring against RBS threats.}, the first comprehensive system integrated with O-RAN for detecting RBS attempting to launch downgrade attacks, combining an IDS xApp with an enhanced telemetry collection mechanism within the Near Real-Time RIC (NearRT-RIC).
It collects diverse physical layer (PHY-layer) indicators, such as Reference Signal Received Power (RSRP), Reference Signal Received Quality (RSRQ), and Signal-to-Interference-plus-Noise Ratio (SINR), extracted from User Equipment (UE) Measurement Reports and Modem Layer 1 data, and leverages Machine Learning (ML) models to detect the presence of malicious cells within the surrounding area.
Furthermore, to address the limitations of restricted real-world measurements, we evaluate our system using an extensive, real-world dataset collected from over 10 areas across two major U.S. cities, covering two MNOs and four Commercial Off-The-Shelf (COTS) UEs.
Finally, to assess the performance of our framework, we built a custom testbed using Open5GS \cite{open5GS}, srsRAN \cite{srsRAN}, and FlexRIC \cite{flexric} as reference platforms for the core network and O-RAN infrastructure.
Our system achieves up to \textit{99.5\% Accuracy} and \textit{96.7\% Precision}, demonstrating both its reliability in detecting rogue cells over time and its robustness against false alarms.
The remainder of the paper is organized as follows:
Section~\ref{sec:related} reviews related work, while Section~\ref{sec:background} provides the necessary background.
Section~\ref{sec:IDS} outlines the threat model and architectural design of the proposed IDS.
Section~\ref{sec:eval} presents the benchmarking and experimental results, whereas Section~\ref{sec:ethical} details the ethical considerations adhered to throughout this work.
Finally, Section~\ref{sec:conclusions} concludes the paper with implications for future work.

\section{Related Work} \label{sec:related}

5G introduces substantial security improvements over previous generations; however, it still inherits vulnerabilities that persist from legacy systems such as LTE \cite{LOGIC_gone},  enabling various attacks, such as IMSI Catching.
IMSI catchers, implemented through RBS, have been widely studied across all cellular generations \cite{SOK, imsi_catch}.
These attacks allow adversaries to actively impersonate legitimate base stations, prompting UEs to reveal their IMSI in plaintext, leading to subscriber identity exposure, tracking, and localization \cite{NAS_Sec, hussain2019privacy, ltrack}. 
Beyond IMSI disclosure, RBS facilitates a range of threats, drawing attention from both academic and standardization communities \cite{SOK}.
The 3rd Generation Partnership Project (3GPP) introduced an optional RBS detection framework within its technical specifications \cite{3gpp33501}, and further dedicated an entire technical report to this issue \cite{3gpp33809}. 
The report identifies critical RBS threat scenarios and introduces mitigation strategies, including enhanced UE-Measurement Reporting.
However, these proposals lack concrete implementation strategies or timelines for integration into the specification.

The ongoing risk posed by RBS is further amplified by the continued reliance on legacy mobile networks like LTE \cite{demestifying}. 
This enables bidding-down attacks, downgrading UEs from 5G to less secure generations, and exploiting weaker authentication and encryption protocols \cite{neverLetmeDown, NewVuln, reasoner}. 
In \cite{neverLetmeDown}, researchers demonstrate a downgrade attack from 5G-SA to 2G on commercial networks.
Similarly, \cite{NewVuln} reveals a vulnerability in LTE where UEs reveal their capabilities before establishing RRC security, allowing adversaries to intercept and manipulate these messages to initiate a downgrade. 

Given the limitations of current defenses, recent work has explored O-RAN as a promising path forward. 
Its architecture has motivated extensive research into leveraging O-RAN as a foundation for IDS \cite{5GSpec, tsourdinis, sdr_rogue_bs, effecient_ids, packet_continuity, det-ran, jamming_oran, malicious_flows, 6G-XSec}, enabling intelligent threat monitoring within the RAN infrastructure.
A prominent example is 5G-SPECTOR \cite{5GSpec}, a framework targeting Layer 3 protocol exploit detection, utilizing a security audit and xApp. 
Similarly, \cite{tsourdinis} presents an AI/ML-driven IDS that also functions as a real-time resource allocator. 
In \cite{sdr_rogue_bs}, the authors propose UE-level detection of RBS by training ML models on signal stability metrics within the NearRT-RIC and distributing them back to the UEs. 
A related effort, \cite{jamming_oran}, focuses on jamming detection using UE-reported Channel Quality Indicator (CQI) and RSRP values, employing the Kolmogorov-Smirnov test to flag anomalies.
\cite{effecient_ids} deploys an IDS within the NearRT-RIC security module, targeting model poisoning attacks in ensemble learning setups. 
\cite{malicious_flows} similarly uses cross-domain AI models embedded in xApps, combining data from both the RAN and transport networks.
Pushing detection to lower layers, Det-RAN \cite{det-ran} proposes a real-time IDS at the gNB-DU, leveraging PHY-layer features such as IQ samples and CSI. 
Meanwhile, \cite{packet_continuity} focuses on the Open Fronthaul (O-FH) interface, applying deep learning to detect and mitigate DDoS attacks.
Finally, 6G-XSec \cite{6G-XSec} introduces a two-stage IDS combining unsupervised anomaly detection via xApp with a Large Language Model (LLM) for threat interpretation. 

Although these studies present insights for IDS, none has explicitly focused on detecting RBS, particularly in the context of downgrade attacks. 
Moreover, they lack actual implementation, relying on simulations and artificial datasets that fail to capture the constraints of real-world mobile networks and UE behavior.
In this work, we address these gaps by designing, implementing, and evaluating a real-time IDS system within O-RAN, utilizing a real-world setup.
The system is evaluated using real-world measurements collected directly from COTS UEs operating on public commercial networks across multiple MNOs.
The system accurately detects malicious cells within a given area, offering a practical implementation that advances ongoing research in RBS detection.

\begin{figure}
    \centering
    \includegraphics[width=1\columnwidth]{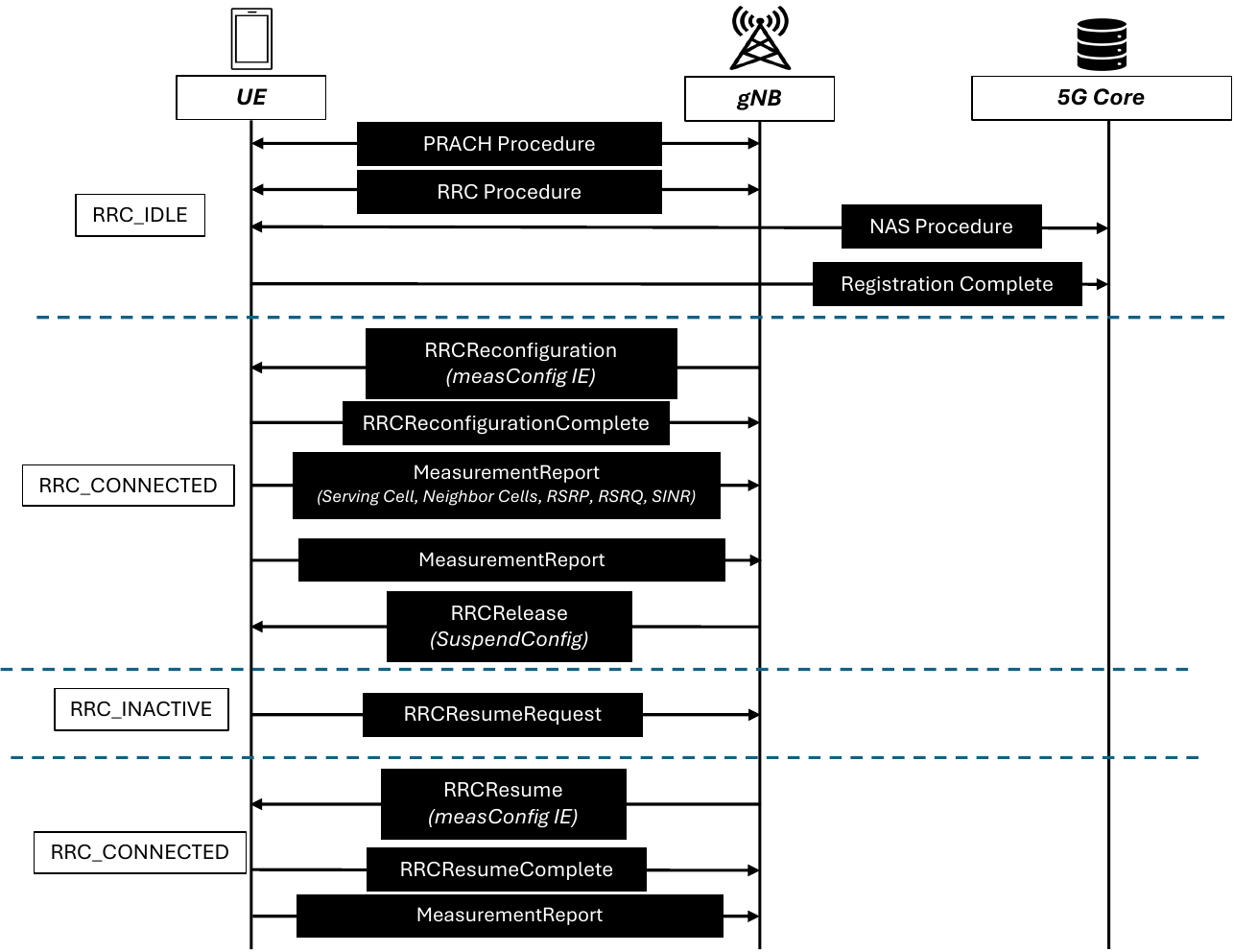}
    \caption{RRC state transitions and Measurement Reports.}
    \label{fig:rrc}
\end{figure}

\section{3GPP/O-RAN Telemetry Mechanisms}
\label{sec:background}

In this section, we present the necessary background, introducing the architecture of LTE/5G networks, the telemetry mechanisms underpinning the Measurement Reporting process, and the key components and interfaces that define O-RAN.

\subsection{Cellular Network Operations}

\subsubsection{5G \& LTE Cellular Networks}


The architecture of 5G systems is composed of three principal entities, as depicted in Figure \ref{fig:rrc}: (1) the UE, typically a smartphone equipped with a Universal Subscriber Identity Module (USIM) subscribing to commercial networks and identified by a unique user identifier known as the Subscription Permanent Identifier (SUPI), referred to as IMSI in LTE and earlier generations; (2) the gNodeB (gNB), the 5G base station operating within the RAN, which connects the UE to the MNO's core network; and (3) the 5G Core Network (5G-CN), a service-based architecture (SBA) enabling authentication, security, and session management through different Network Functions (NFs).
The gNB may interface either with a 5G Core Network (5G-CN) in the 5G SA architecture or with an LTE Evolved Packet Core (EPC) in the 5G Non-Standalone (NSA) architecture.

The initial procedure for UE connectivity begins with cell attachment and network registration. 
The UE performs initial cell selection by detecting and decoding System Information Block (SIB) messages broadcast by nearby gNBs. 
Subsequently, it initiates random access via the Physical Random Access Channel (PRACH) to achieve uplink synchronization. 
Upon successful random access, the establishment, maintenance, and release of radio connections are managed by the Radio Resource Control (RRC) protocol \cite{3gpp38331}.
To establish an RRC connection, the UE sends an RRCSetupRequest message. 
If the gNB accepts, it responds with an RRCSetup message providing configuration information. 
The handshake is then completed with an RRCSetupComplete message.
Following the RRC establishment, the Non-Access Stratum (NAS) procedures commence to facilitate UE registration with the core network. 
The UE exchanges NAS messages with either the Access and Mobility Management Function (AMF) in the 5G-CN or the Mobility Management Entity (MME) in the EPC.
The NAS procedure is initiated with a Registration Request (in 5G) or Attach Request (in LTE), containing the UE’s temporary (TMSI) or permanent identifiers, such as the Subscription Concealed Identifier (SUCI) in 5G or the IMSI in earlier generations.
Authentication and security procedures follow, involving the Authentication and Key Agreement (AKA) protocol. 
Upon successful authentication, the NAS procedure concludes with a Registration Complete or Attach Complete message.

Once a secure radio connection between the UE and the network has been established, the UE enters the RRC\_Connected state. 
In this state, the network can transmit system information to the UE through dedicated signaling, primarily using the \textit{RRCReconfiguration} message. 
This procedure is used to modify an already established RRC connection by configuring various parameters \cite{3gpp38331}.
Once the UE successfully acknowledges the reconfiguration process, it responds with an RRCReconfigurationComplete message to confirm the changes.
If the network is in an idle state, the UE can optionally stay in RRC\_Inactive state, as depicted in Figure \ref{fig:rrc}, instead of completely releasing the RRC connection and recover it via the \textit{RRC\_Resume} procedure.
During the RRC\_Resume procedure, the exchange of messages such as RRCResumeRequest, RRCResume, and RRCResumeComplete occurs.


\subsubsection{UE Measurement Reports}

Following the successful establishment of the RRC connection, the network can instruct the UE to perform specific measurements through the \textit{measConfig} Information Element (IE), typically conveyed within the RRCReconfiguration \cite{3gpp38331}, as shown in Figure \ref{fig:rrc}.
This IE defines the measurement configuration that the UE should follow, specifying which frequencies, cells, or signals to monitor. 
measConfig IE may also be included within the RRCResume message. 
The measurement configuration can direct the UE to perform intra- and inter-frequency NR measurements, defined through the \textit{MeasObjectNR} IE, as well as inter-Radio Access Technology (RAT) measurements, including E-UTRA (LTE) measurements via the \textit{MeasObjectEUTRA} IE and UTRA-FDD (UMTS) measurements via the \textit{MeasObjectUTRA-FDD} IE. 
Depending on the configuration, measurements can be based either on Synchronization Signal/Physical Broadcast Channel blocks (SSB/PBCH) or on Channel State Information Reference Signals (CSI-RS).
Measurement Reporting can be configured to occur periodically, via the \textit{reportInterval} IE
or based on event-triggered conditions.
Additionally, the network may provide lists of specific cells that the UE should prioritize or cells that should be ignored.
Furthermore, the network specifies the radio quantities to be included in the reports, such as RSRP, RSRQ, or SINR. 
Through these configurations, Measurement Reporting enables the UE to provide the network with critical information regarding the radio environment, supporting functions such as mobility management, beam selection, handover, and connection optimization.


\begin{figure}
    \centering
    \includegraphics[width=1\columnwidth]{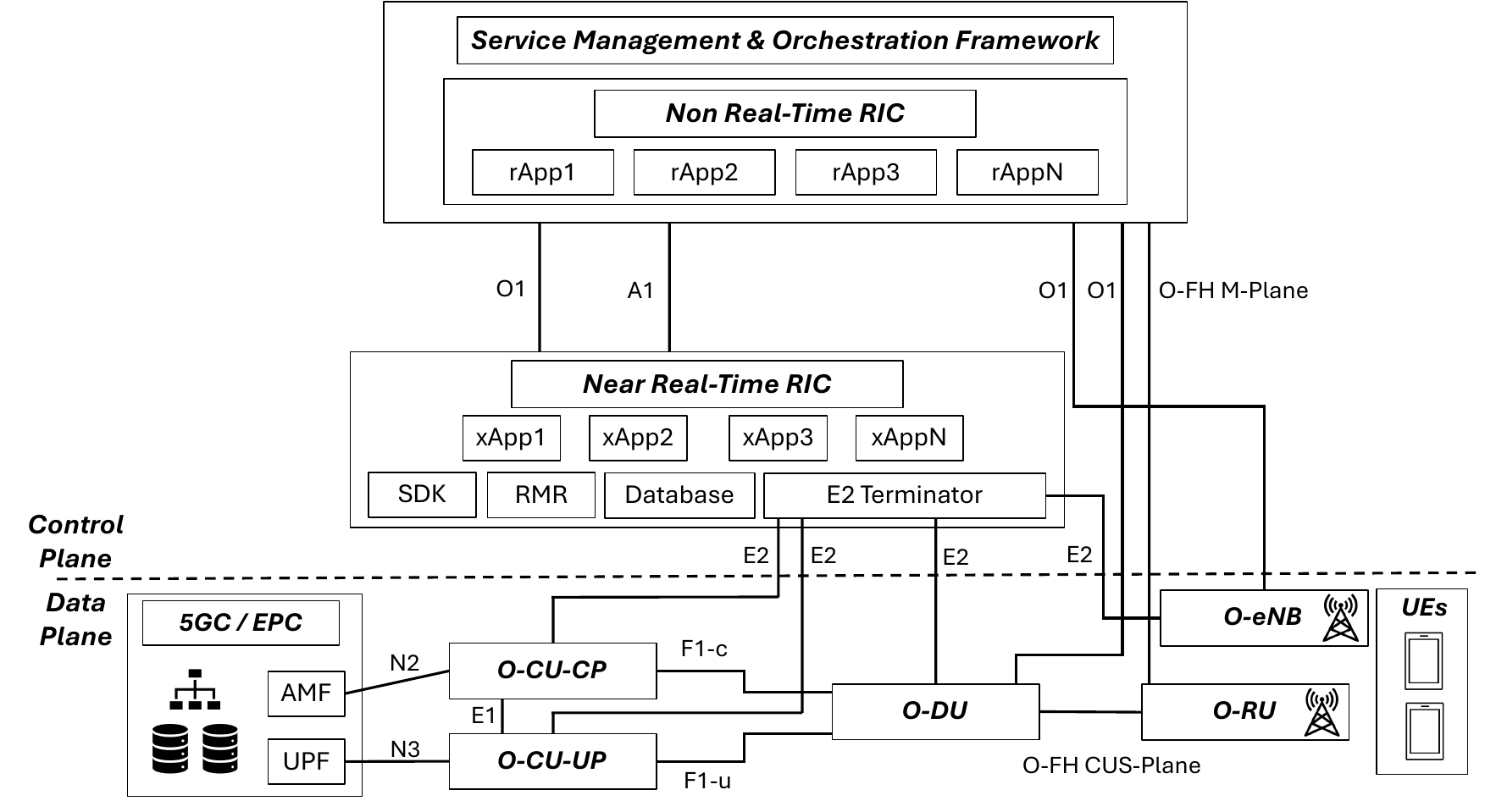}
    \caption{Open Radio Access Network (O-RAN) architecture.}
    \label{fig:o-ran}
\end{figure}


\begin{figure*}
    \centering
    \includegraphics[width=2\columnwidth]{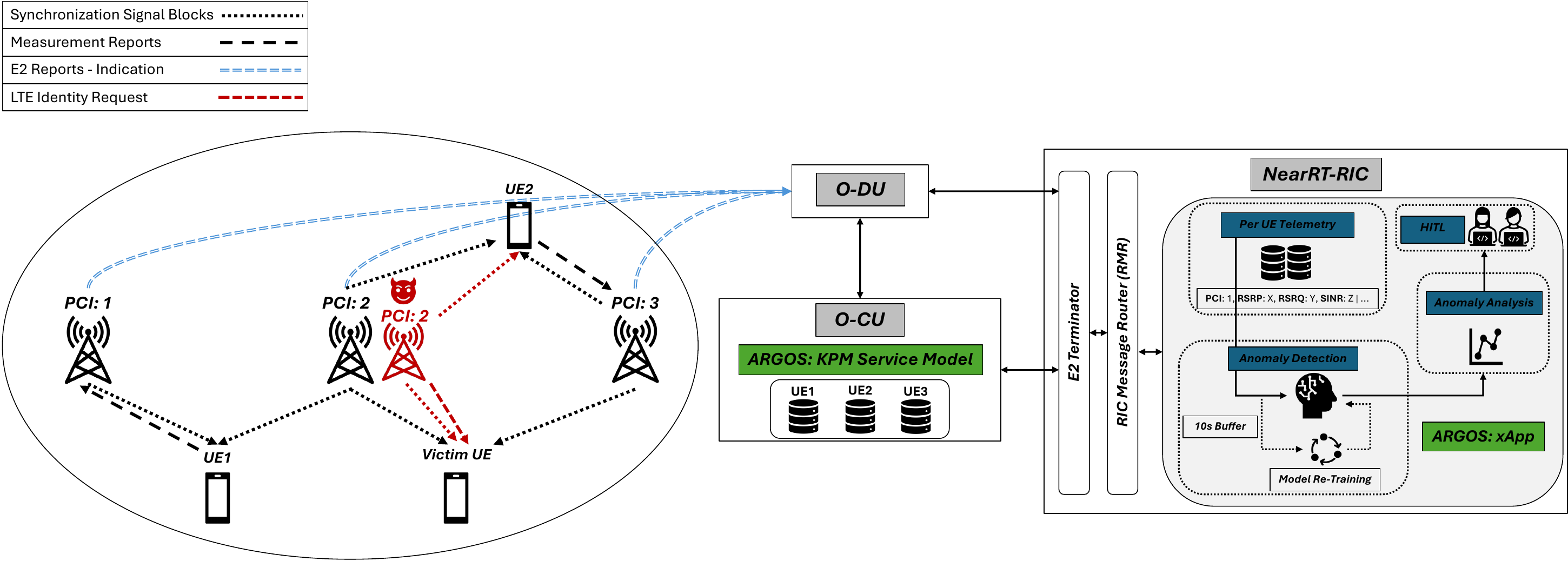}
    \caption{Even if an RBS mimics the PCI of a legitimate cell, it cannot replicate the RF propagation profile observed by all UEs. 
    These measurements are transmitted to the Near-RT RIC via \system’s enhanced KPM Service Model, where a dedicated xApp leverages trained ML models to detect anomalies.}
    \label{fig:malicious}
\end{figure*}

\subsection{O-RAN Architecture}

Figure~\ref{fig:o-ran} illustrates the O-RAN architecture, highlighting both the data and control planes.
Below, we outline the key architectural principles governing each plane.

\subsubsection{O-RAN Data Plane}

The O-RAN architecture, illustrated in Figure~\ref{fig:o-ran}, follows the 3GPP disaggregation model \cite{3gpp38401}, splitting the gNB into three main units: the Radio Unit (O-RU), Distributed Unit (O-DU), and Central Unit (O-CU).
O-RUs, located near the antennas in the fronthaul, handle PHY-layer operations.
O-DUs and O-CUs, deployed at the network edge, manage Layers 2 and 3.
The O-DU oversees Medium Access Control (MAC) and Radio Link Control (RLC) functions, while the O-CU, divided into O-CU-CP (control plane) and O-CU-UP (user plane), handles RRC and forwards control/user traffic to the core network (AMF/UPF).
Standardized interfaces connect these components, with F1 linking O-DU and O-CU and E1 connecting O-CU-CP to O-CU-UP.


\subsubsection{O-RAN Control Plane}

The O-RAN control plane is distinct from the data plane and centers on the RICs, which are split into NearRT-RIC and Non Real-Time RIC (NonRT-RIC). 
These programmable components provide centralized network visibility, enabling closed-loop control and orchestrating RAN operations.
The NearRT-RIC operates on timescales from 10 milliseconds to 1 second and hosts modular control applications called xApps for different use cases.
Communication between xApps and other internal components is facilitated by the RIC Message Router (RMR), which ensures efficient message routing within the NearRT-RIC.
NearRT-RIC interfaces directly with the O-DU and O-CU, also referred to as E2 Nodes, via the E2 interface, allowing real-time telemetry and control. 
Interactions are governed by four core E2 procedures: \textit{Report, Insert, Control, and Policy}. 
Each xApp implements specific E2 Service Models (E2SMs) layered over the E2 Application Protocol (E2AP).
The NonRT-RIC, part of the Service Management and Orchestration (SMO) framework, operates on longer timescales (above 1 second), supporting similar control applications to xApps, known as rApps, which are higher-level applications that generate long-term policies impacting overall network behavior.


\section{\system\ Overview}
\label{sec:IDS}

In this section we outline the threat model targeted by our system and present an architectural overview of \system, detailing UE telemetry acquisition and employed ML models.

\subsection{Exploiting LTE Compatibility in COTS UEs}

The threat model addressed in this work concerns the exploitation of plaintext IMSI transmission over the air. 
As discussed in Section~\ref{sec:background}, during the initial NAS message exchange between the UE and the network, the UE transmits either a permanent or temporary identifier. 
In the absence of prior interaction or under malicious intent, the network may issue an Identity Request, prompting the UE to disclose its permanent identifier. 
In LTE networks specifically, the IMSI is transmitted without encryption or integrity protection, enabling adversaries to intercept it.
Such behavior enables IMSI catching attacks, in which RBS masquerade as legitimate cells, broadcasting identical network identifiers with stronger signal or with a different Tracking Area Code (TAC), deceiving the UE into believing that it has entered a new tracking area.
RBS can either broadcast a different Physical Cell ID (PCI) than nearby legitimate cells or carry out a more sophisticated attack by reusing the same PCI to impersonate a valid cell, as illustrated in Figure~\ref{fig:malicious}.
Once camped at RBS, the UE is coerced to disclose its IMSI, or in some cases even its IMEI~\cite{neverLetmeDown}, enabling subsequent user tracking and localization~\cite{SOK}.
This vulnerability stems from specifications in the 3GPP standard rather than implementation flaws, rendering all LTE-compatible devices susceptible regardless of vendor. 
The issue is expected to persist until UEs fully transition to support only the latest standards, disabling previous generations.
As highlighted in~\cite{ericsson_mobility}, current deployment limitations of 5G-SA make it infeasible to release UEs that are exclusively 5G-compatible. 
Therefore, coordinated efforts between UE manufacturers and telecom vendors are essential to promote end-to-end adoption of secure, modern standards. 
In the interim, we propose \system\  as a practical and deployable solution to detect and mitigate RBS effectively within existing network environments.

\begin{algorithm}
\vspace*{0.05in}
\caption{\system\ Telemetry-Based Anomaly Detection}
\label{alg:telemetry}
\begin{algorithmic}[1]

\State \textbf{Input:} $\mathrm{T}_u = \{([\mathbf{f}_u], [\mathbf{c}_u], [\mathbf{r}_u], [\mathbf{q}_u], [\mathbf{s}_u], [\mathbf{t}_u])\}$ for each UE $u = 1, \ldots, N$; ML model $\mathbf{M}$; Anomaly threshold $\tau$

\State \textbf{Output:} Anomaly score per second $\alpha_u(t)$; MSE per $u$
\State Connect to Near-RT RIC.
\State Subscribe to E2 nodes using KPM Service Model.
\For{$u$}
        \State Circular Buffer $\mathrm{B}_u \gets \emptyset$
\EndFor

\While{true}

\For{$u$}
    \State $\mathrm{B}_u  \gets \mathrm{T}_u$
    \If{$\mathrm{B}_u$ $>=$ 1 second of new telemetry}
         \State $X_u(t) \gets encoded(\mathrm{B}_u)$ 
         \State $\hat{X}_u(t) \leftarrow
         \mathbf{M}(X_u(t))$
        \State $\alpha_u(t) \leftarrow \text{MSE}(X_u(t), \hat{X}_u(t))$
        
        \If{$\alpha_u(t) > \tau$}
        \State {Anomaly}
        \Else  \State{Legitimate}
        \EndIf
    \EndIf
\EndFor

\If{$\sum_{u=1}^{N} |\mathrm{B}_u| \geq 10 \text{ seconds}$}
    \State $\mathrm{D} \leftarrow \bigcup_{u=1}^{N} \mathrm{B}_u$
    \State $\mathbf{M} \leftarrow \text{Train}(\mathrm{D})$
    \State $\tau \leftarrow \text{GetThreshold}(\mathbf{M}, \mathrm{D})$
\EndIf


\EndWhile
\end{algorithmic}
\end{algorithm}

\subsection{\system\ Architecture}

The E2 Setup procedure serves as the starting point of the architecture and is independent of any specific xApp.
This procedure establishes application-level communication between E2 Nodes and the NearRT-RIC, replacing any prior configurations with the latest agreed-upon parameters \cite{oran.e2ap}. 
Importantly, during this phase, E2 Nodes expose the types of telemetry and control capabilities they support.
Following successful setup, xApps can query information about connected E2 Nodes and initiate telemetry collection via the E2 Subscription procedure \cite{oran.ricarch.v7}, as shown in Figure~\ref{fig:malicious}. \system\ uses only the \textit{Report} procedure, which involves \textit{E2 RIC Indication} messages containing telemetry from E2 nodes, sent either periodically or in response to specific trigger events.
However, the system is extensible to support the remaining services introduced in Section~\ref{sec:background}, enabling a shift from IDS to IPS.

To collect telemetry at the xApp level from the E2 nodes, we adopt the latest Key Performance Measurement (KPM) Service Model defined by 3GPP \cite{3gpp28552}. 
Several KPMs defined in the latest Service Model encapsulate Measurement Report messages from UEs, which, based on the measConfig IE, include metrics such as RSRP, RSRQ, and SINR.
This Model is extended by \system, incorporating both intra- and inter-frequency measurements (5G and LTE), enabling a comprehensive view of all neighboring cells as reported directly by UEs.

Each E2 Node, particularly the O-CU handling RRC signaling, aggregates Measurement Reports per UE, identified by SUPI or SUCI, with a dedicated memory buffer assigned to each UE.
Similarly, the implemented xApp maintains its own per UE memory buffers to enable continuous telemetry processing.
Upon receipt at the O-CU, reports are parsed to extract per cell RSRP, RSRQ and SINR measurements. 
After one second of telemetry is accumulated, the data are encapsulated in E2 RIC Indication messages, structured according to the extended KPM Service Model, and sent to the xApp via the E2 interface, where anomaly detection is performed using deep learning techniques, as depicted in Figure \ref{fig:malicious}.
Before being passed to the models for evaluation, the received telemetry undergoes additional processing to generate per-second vectors.
As shown in Algorithm~\ref{alg:telemetry}, each per-second vector $\mathrm{T}_u$ encodes the presence or absence of known legitimate cells, identified by their Absolute Radio Frequency Channel Number (ARFCN) ($\mathrm{f}_u$) and PCI ($\mathrm{c}_u$).
It includes corresponding measurements of RSRP ($\mathrm{r}_u$), RSRQ ($\mathrm{q}_u$), and SINR ($\mathrm{s}_u$), each independently normalized. The vector also records the measurement timestamp associated with each observation.
During inference, the xApp evaluates each per-second vector, generating a binary anomaly verdict along with the associated Mean Squared Error (MSE) value.
To adapt to evolving network behavior, the model is retrained every 10 seconds using newly accumulated legitimate telemetry. 
The telemetry collection and preprocessing workflow is summarized in Algorithm~\ref{alg:telemetry}, along with the model's inference and re-training processes.

\subsection{Deep Learning Based RBS Detection}

To ensure safe integration with commercial networks and adhere to ethical standards, outlined in Sections~\ref{sec:eval} and~\ref{sec:ethical}, \system\ is trained and tested exclusively on legitimate, non-malicious data collected from commercial MNOs using passive observation setups.
Given the absence of labeled attack data, our xApp is evaluated using four unsupervised learning models: Autoencoders, Denoising Autoencoders, Variational Autoencoders, and Isolation Forests. 
These models are inherently suited for anomaly detection tasks where only benign patterns are available during training.

\subsubsection{\textit{Autoencoders}}

Autoencoders (AEs) are artificial neural networks designed to learn compact representations of unlabeled data.
Their architecture includes an encoding function that compresses the input vector into a lower-dimensional space, $A: \mathbb{R}^n \rightarrow \mathbb{R}^p$, and a decoding function that reconstructs the original vector $A: \mathbb{R}^p \rightarrow \mathbb{R}^n$ \cite{ae}.
Together, these functions aim to minimize the reconstruction error, computed using the MSE loss in Equation~\ref{eq:MSE}, and optimized via backpropagation to capture the input data distribution.
Vectors with high MSE values are flagged as anomalous, indicating potential outliers.
In \system, the anomaly threshold is set after training using the 99.9\textsuperscript{th} percentile of MSE values from the training dataset.
This approach minimizes the likelihood of legitimate vectors being incorrectly flagged as anomalous (False Positives), while maintaining sensitivity to suspicious patterns.


\begin{equation}
\mathcal{L}_{\text{MSE}} = \frac{1}{n} \sum_{i=1}^{n} \left( x_i - \hat{x}_i \right)^2
\label{eq:MSE}
\end{equation}

AEs are well-suited for the discussed problem, based on their ability to learn the underlying patterns of per-second measurements, capturing typical combinations of cells and their associated signal characteristics. 
By reconstructing these vectors, the AE effectively models normal telemetry behavior, enabling the detection of deviations indicative of anomalies. 
Given the presence of measurement noise due to reflections and other propagation effects, we extend the baseline AE to include a Denoising AE and a Variational AE, which improve generalization and mitigate overfitting by learning robust latent representations.

\subsubsection{\textit{Denoising-Autoencoder}}

Denoising Autoencoders (DAEs) are more robust variants of AEs, used for error correction.
In \system, the DAE shares the same architecture as the standard AE, with the key difference being that input training vectors are corrupted with Gaussian noise. 
The model is then trained to reconstruct the original, noise-free vectors.
The noise process is modeled by a function $T: X \rightarrow X$, where $T(x) = x + \epsilon$ and $\epsilon$ is sampled from a Gaussian distribution $\mu_T = \mathcal{N}(0, \sigma^2)$.
This method assists the network in avoiding the memorization of the input,  forcing it to learn the core features of the dataset.

\subsubsection{\textit{Variational-Autoencoder}}

Similar to DAEs, Variational Autoencoders (VAEs) share the same architecture as standard AEs but are grounded in the mathematical framework of Variational Bayesian (VB) methods.
In VAEs, the encoder maps each input vector to a Gaussian distribution in the latent space, parameterized by a mean vector $\mu$ and a standard deviation vector $\sigma$.
A latent vector is then sampled from this distribution, and the decoder attempts to reconstruct the original input.
The loss function combines a reconstruction loss (MSE) and a Kullback–Leibler (KL) divergence loss, as shown in Equation~\ref{eq:kl_continuous}, which regularizes the latent space by encouraging it to match a prior distribution.
This probabilistic formulation helps prevent overfitting and improves generalization.

\begin{equation}
D_{\mathrm{KL}}(P \,\|\, Q) = \int_{-\infty}^{\infty} P(x) \log \left( \frac{P(x)}{Q(x)} \right) dx
\label{eq:kl_continuous}
\end{equation}

\subsubsection{\textit{Isolation Forest}}

Isolation Forests are a well-established anomaly detection algorithm based on binary trees.
The core idea is that anomalies, being few and different, can be isolated with fewer partitions.
The algorithm recursively builds \textit{Isolation Trees} by randomly selecting an attribute and a split value between its minimum and maximum range.
Anomaly scores are derived from the path length, as anomalies typically require fewer splits to be isolated.
However, Isolation Forests assume anomalies are few and different in feature space, something that may limit performance in datasets with subtle or high-density anomalies.

\section{Experimental Evaluation}
\label{sec:eval}

This section presents the experimental evaluation of the proposed framework, detailing the deployed software and hardware components, along with a performance analysis of \system\ from both system and ML model perspectives.

\subsection{\system\ O-RAN Compliant Testbed}

To evaluate the proposed system, a controlled O-RAN-compliant testbed,  including 5G SA, LTE, 5G-CN/EPC and NearRT-RIC components, is deployed. 
All software-based components are deployed within the same x86\_64 Ubuntu 22.04.4 LTS host, equipped with 8 11th Gen Intel Core i7-1195G7 @ 2.90 GHz, 32.0 GiB RAM and 1.0 TB disk capacity.
The testbed leverages version 24.10 of srsRAN, version 23.11 of srsUE, latest version of Open5GS Release-17, and latest version of the br-flexric branch of FlexRIC to emulate real-world gNB (5G-RAN), eNB (LTE RAN), 5G-CN/EPC and NearRT-RIC behavior accordingly. 
The RU front end of the deployed networks is hosted within 2 Ettus Research Universal Software Radio Peripheral (USRP) X310 SDR devices. 
One Pixel 5 COTS UE is utilized, equipped with a sysmocom SIM card programmed with PLMN identifiers matching the 5G SA deployment.
The core component of \system, our ML-based xApp, is integrated into the NearRT-RIC through FlexRIC, supporting both Python and C implementations.
Communication between CU and NearRT-RIC is established over the standardized E2 interface, while FlexRIC’s internal E42 interface facilitates communication between xApp and the RIC controller.

The testbed, configured using srsRAN-provided files \cite{srs_Hand}, is used to demonstrate the feasibility of the RBS Downgrade attack.
This setup enables active UE Measurement Reporting and supports handover, facilitating inter-cell movement closely replicating real-world deployments.
Once the adversarial eNB becomes active, it impersonates the legitimate network’s PLMN ID as well as PCI while transmitting at a higher signal strength.
The UE connects to the adversarial eNB and exposes its IMSI, resulting in the successful compromise of its identity, demonstrating both the feasibility and simplicity of the attack.

\begin{table}
\vspace*{0.05in}    
\centering
\vspace*{0.05in}
\begin{tabular}{ll}
\toprule
\textbf{Metric} & \textbf{Equation} \\
\midrule
Accuracy & $(TP + TN) / (TP + FP + TN + FN)$ \\
Precision & $TP / (TP + FP)$ \\
Recall & $TP / (TP + FN)$ \\
F1-Score & $2 \cdot \text{Precision} \cdot \text{Recall} / (\text{Precision} + \text{Recall})$ \\
\bottomrule
\end{tabular}
\vspace{0.5em}
\caption{Performance Metrics Used for ML Model Evaluation.}
\label{tab:ml_metrics}
\vspace{-1.5em}
\end{table}

\subsection{Real-World Data Collection}

To ensure realism beyond controlled testbed conditions, the system is evaluated using real-world data collected directly from commercial networks operated by different MNOs.
Over a three-month period in 2025, 5G and LTE measurements were collected at various times across 10 urban areas in Boston and San Francisco.
The dataset covers two major U.S. MNOs and was curated to capture real-world behavior of the Measurement Reporting mechanisms. 
Data collection was performed using rooted COTS UEs, including two Google Pixel 5 devices, an LG Velvet 5G, and a OnePlus 8 5G, all equipped with measurement tools such as Network Signal Guru (NSG) \cite{nsg}. 
Devices were carried in motion through the areas while connected to commercial networks.
Logged data was later analyzed using Qualcomm's QXDM \cite{qxdm} to extract low-level Modem and Layer 1 metrics. 
In total, our dataset comprises 22,626 seconds of telemetry and 232,810 NSG-QXDM data points, with each point capturing detailed per-cell measurements, including PCI, RSRP, RSRQ, and SINR of neighboring cells.

\begin{figure}[t]
    \centering
    \includegraphics[width=1\columnwidth]{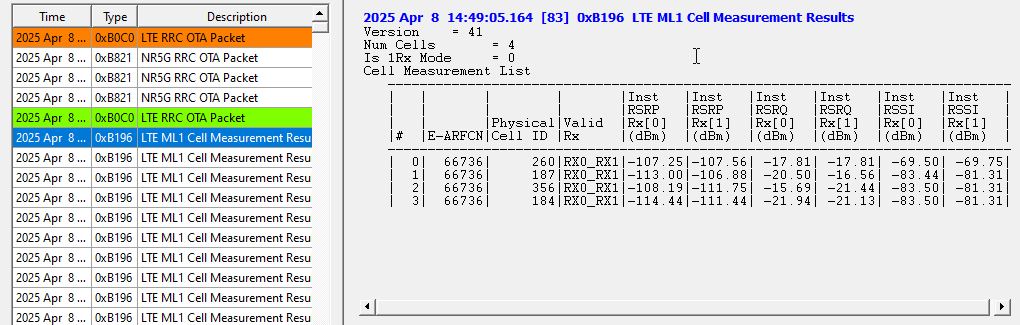}
    \caption{UE Modem Layer 1 (ML1) Cell Measurements captured via QXDM.}
    \label{fig:ml1}
\end{figure}

In addition to standard Measurement Reports, we also incorporate \textit{Modem Layer 1 (ML1)} Cell Measurement Results, obtained via QXDM, as depicted in Figure~\ref{fig:ml1}. 
Unlike traditional Measurement Reports, ML1 data provide high-frequency sampling of neighboring cell signal stability, offering a much finer temporal resolution.
ML1 data are parsed similarly to standard Measurement Reports, as they share the same underlying structure.
As a result, the telemetry vectors used by \system\ maintain a consistent format while being enriched with additional measurements, thereby enhancing the dataset’s granularity.
However, the main limitation of ML1 data is that they are never transmitted to the gNB over the air (OTA) and remain accessible only at the UE side.
Since operator-configured Measurement Reports are often sparse or event-triggered, ML1 data significantly enhance the observability of UE behavior by providing a more continuous and detailed stream of measurements.
We advocate for the integration of ML1 data into O-RAN control loops to support more accurate and timely anomaly detection.

\subsection{Rogue Cell Inclusion}

The performance of \system\ is evaluated using both benign and malicious cellular network traffic.
Since the dataset contains only legitimate data, two types of RBS strategies, \textit{Adversary 1 (A1)} and \textit{Adversary 2 (A2)}, are emulated by selecting a valid cell, excluding it during training, and reintroducing it during inference.
This methodology allows for realistic evaluation of our system by leveraging authentic cell behavior, eliminating the need for synthetic data generation.
A1 does not replicate the PCI of an existing cell. 
As a result, the reintroduced cell during inference retains its identity, allowing us to assess the model’s ability to detect previously unrecognized cells.
A2 carries out a more intricate attack by replicating the PCI of a legitimate cell, causing the reintroduced cell during inference to share the same PCI as an existing legitimate one.
In the case of A2, the replicated PCI reappears across multiple per-second telemetry vectors where it was previously absent, creating unusual cell combinations.
The model is evaluated both on abnormal co-occurrence patterns in Measurement Reports and on its ability to detect anomalies based on signal characteristics, as A2 instances, despite sharing the same PCI, will exhibit at least slight differences in power, relatively to learned power-levels in conjunction with neighboring cells.



\subsection{Performance Evaluation}

To evaluate our solution, we first compare the performance of the ML models integrated into \system, followed by an assessment of their system-level impact on the RIC platform.

\begin{figure}[ht]
    \centering
    \subfigure[Per-Second Rogue Cell Count $\geq 2$.]{%
        \includegraphics[width=0.47\textwidth]{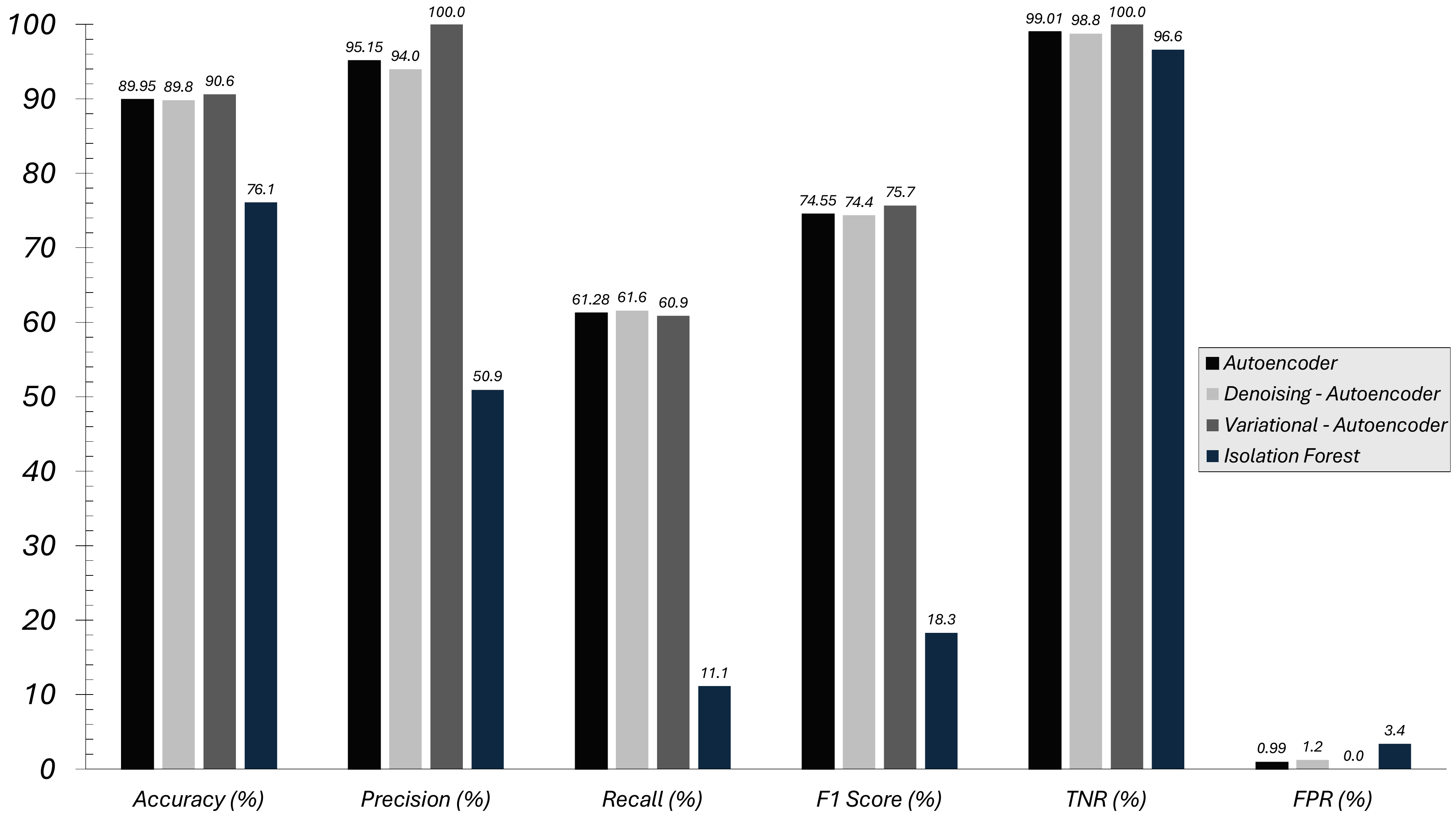}
        \label{fig:ml_performance_2}
    }
    \hfill
    \subfigure[Per-Second Rogue Cell Count $\geq 3$.]{%
        \includegraphics[width=0.47\textwidth]{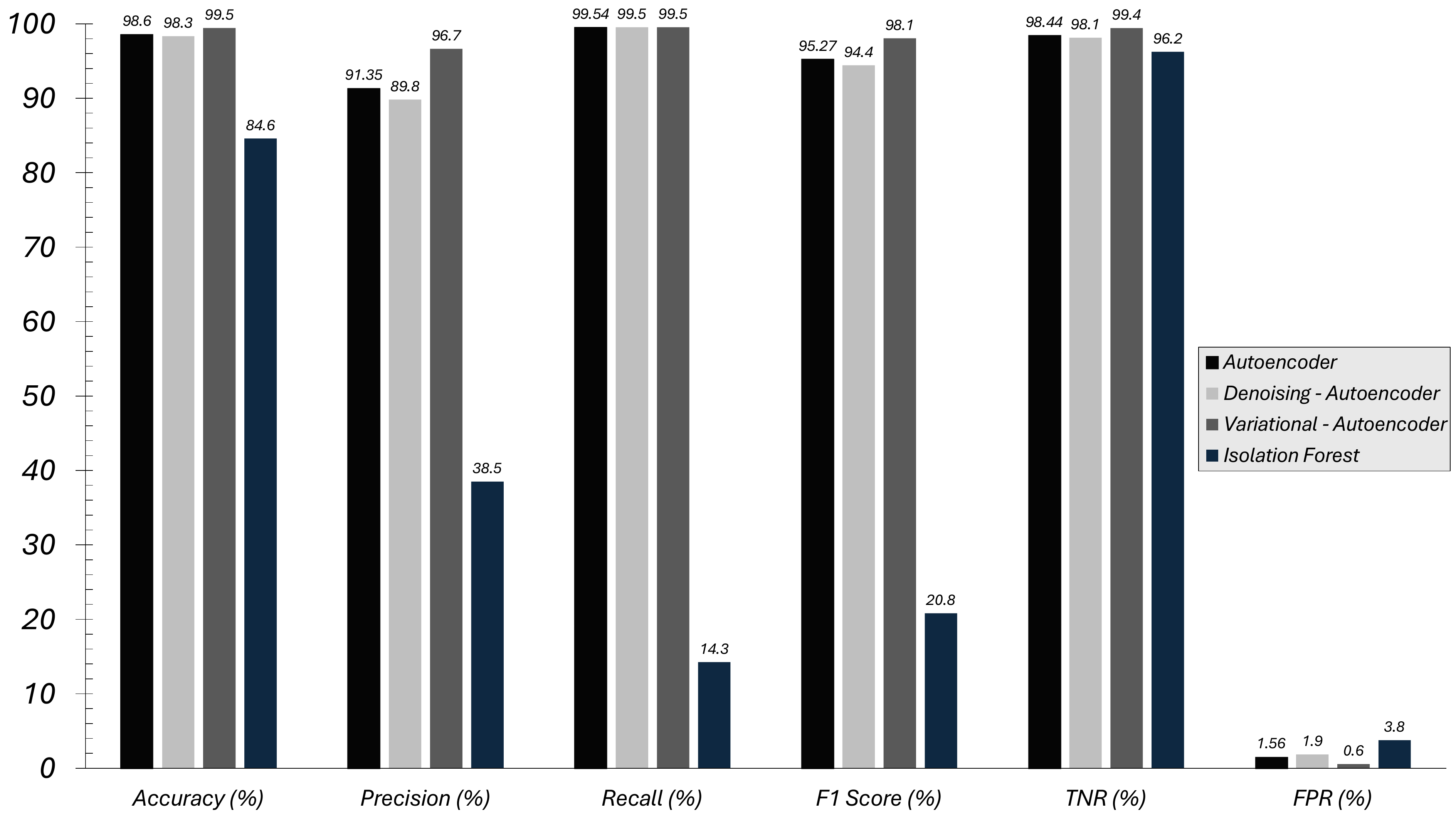}
        \label{fig:ml_performance_3}
    }

    \vspace{1em} 

    \subfigure[Per-Second Rogue Cell Count $\geq 4$.]{%
        \includegraphics[width=00.47\textwidth]{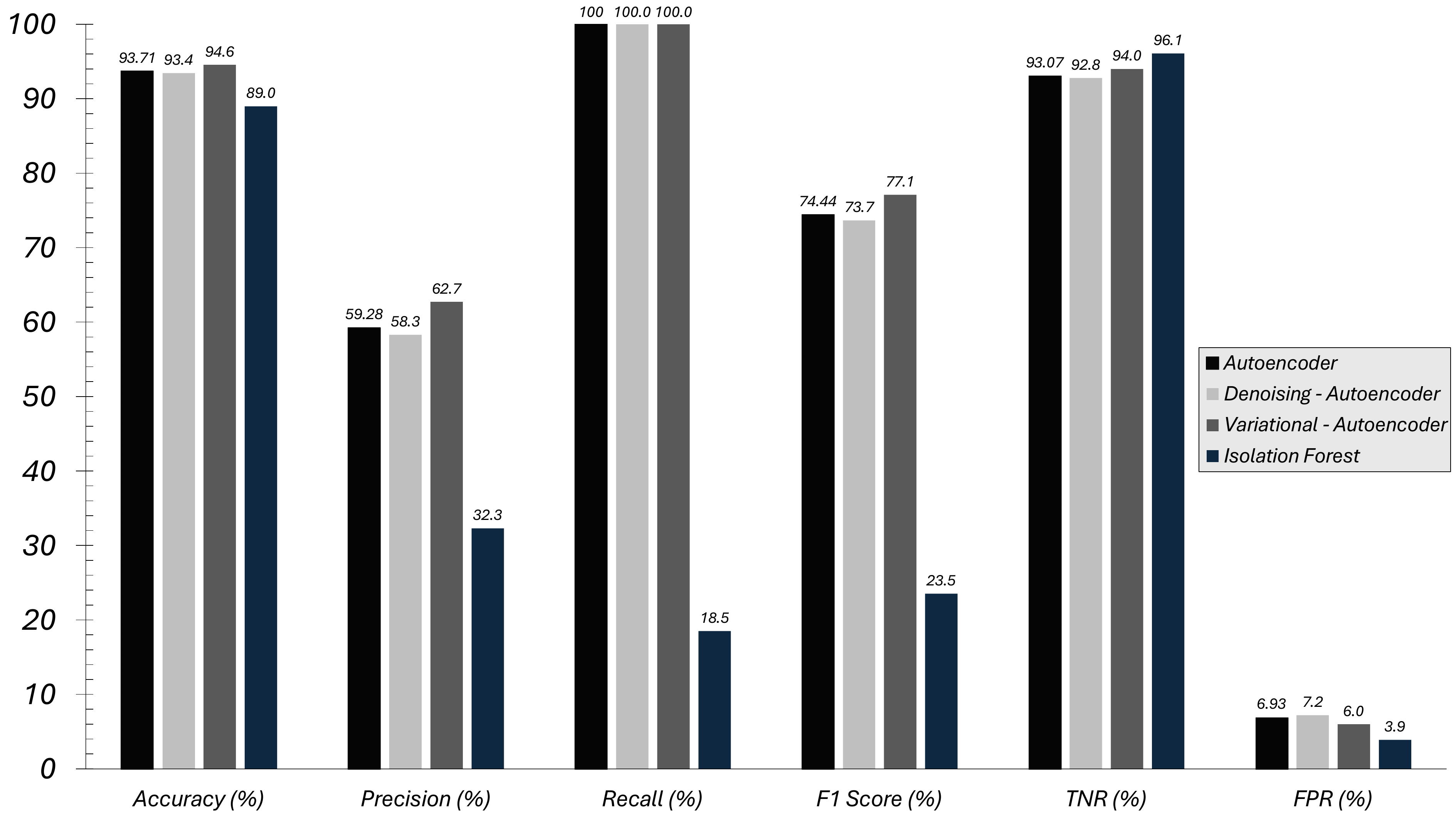}
        \label{fig:ml_performance_4}
    }

    \caption{ML model inference performance across rogue cell appearance thresholds.}
    \label{fig:ml_performance_all}
\end{figure}

To assess the performance of the ML models, we use four standard classification metrics defined in Table \ref{tab:ml_metrics}.
In Table \ref{tab:ml_metrics}, the value TP stands for True Positives, TN for True Negatives, FP for False Positives and FN for False Negatives.
Accuracy provides an overall correctness measure, while Precision emphasizes the proportion of true anomalies among all flagged instances. 
Recall captures the model’s ability to detect all actual anomalies, and the F1-Score provides the harmonic mean between Recall and Precision, especially valuable for imbalanced datasets.
It is important to mention at this stage that, under realistic conditions, if a rogue cell exists within a particular vicinity, regardless of its PCI or signal characteristics, the UE ML1 and Measurement Reports would reflect its presence with high frequency every second, and consequently, so would the per-second telemetry vectors sent to the xApp.
As a result, we consider as anomalous those per-second vectors in which the reintroduced rogue cell appears more than a certain number of times, while the remaining vectors are considered legitimate.
More specifically, as shown in Figure~\ref{fig:ml_performance_all}, we evaluate the ML models in terms of anomaly detection for seconds where the reintroduced cell appears at least 2, 3, or 4 times, a threshold we define as Per-Second Rogue Cell Count.

It is evident that across all AE variations, the best performance is achieved when the Per-Second Rogue Cell Count is $\geq 3$. 
As shown in Figure~\ref{fig:ml_performance_all}, the VAE attains the highest performance under this condition, reaching 99.5\% Accuracy, 97.7\% Precision, 99.5\% Recall, and a 98.1\% F1 Score, with a False Positive Rate (FPR) as low as 0.6\%. 
Both the AE and DAE also demonstrate strong performance, achieving 98.6\% and 98.3\% Accuracy, respectively, while maintaining FPR values below 1.9\%. 
In contrast, the Isolation Forest underperforms across all evaluation metrics, with a maximum Accuracy of 84.6\%, indicating that the randomized attribute selection is suboptimal for capturing the temporal patterns in Measurement Report behavior.

For a Per-Second Rogue Cell Count threshold of $\geq 2$, the models achieve their highest Precision, reaching 100\% for the VAE.
However, the performance of the remaining metrics declines, indicating that while the models are highly effective at avoiding false alarms, they struggle to correctly classify seconds with sparse rogue cell presence.
This results in a drop in Accuracy, as such vectors are often misclassified as legitimate.
Conversely, for a threshold of $\geq 4$, Recall reaches its peak, with all illegitimate seconds correctly identified as anomalous, but at the cost of a higher FPR.
Based on this analysis, a threshold of $\geq 3$ offers the most balanced performance, maintaining both high Accuracy and low FPR. 
For specific operational goals, the system can be configured with alternative thresholds, depending on the desired trade-off between Precision and Recall.

Additionally, we evaluate the impact of \system\ on the control plane, specifically focusing on the NearRT-RIC. 
To this end, we assess the training and inference times, as well as the CPU and memory overhead, across all implemented ML models, as presented in Table~\ref{tab:sys_perf}. 
The evaluation is performed using input datasets of 2000 seconds for training and 500 seconds for inference, drawn from the same area, and executed using a single CPU core on the host machine.
As shown in Table~\ref{tab:sys_perf}, the VAE achieves the lowest training (81.99 seconds) and inference (0.27 seconds) times among all AE variants, rendering it suitable for real-time systems. 
In contrast, the Isolation Forest yields the fastest training time (0.44 seconds) but the highest inference time (11.59 seconds), shifting its computational burden to inference.
Regarding CPU utilization, all AEs occupy approximately 65–68\% of a single core, whereas the Isolation Forest reaches up to 99.16\%, accounting for its prolonged inference duration. 
Lastly, in terms of memory overhead, the Isolation Forest exhibits the highest usage at 548 MB, only slightly exceeding that of the AEs, which peak at 528 MB.

\begin{table}
\vspace*{0.05in}
\centering
\vspace*{0.05in}
\begin{tabular}{lllll}
\toprule
\textbf{Model} &
\textbf{Train(s)} & \textbf{Infer(s)} & \textbf{CPU(\%)} & \textbf{Memory(MB)}  \\
\midrule
AE & 109.50 & 0.28 & 65.20 & 527.20 \\
DAE & 158.68 & 0.28 & 67.16 & 527.24 \\
VAE & 81.99 & 0.27 & 67.25 & 528.72 \\
IF & 0.44 & 11.59 & 99.16 & 548.23 \\
\bottomrule
\end{tabular}
\vspace{0.5em}
\caption{System performance of xApp ML models in the Near-RT RIC.}
\label{tab:sys_perf}
\vspace{-1.5em}
\end{table}

\section{Ethical Considerations}
\label{sec:ethical}

Due to ethical considerations and applicable legal frameworks, it is essential to clarify the methodology used in both our isolated testbed experiments and real-world measurements. 
All active RBS attack scenarios involving over-the-air transmissions were conducted exclusively within our isolated testbed environment, ensuring no interference with operational commercial networks. Specifically, all RF transmissions from srsRAN base stations and UEs were confined to a shielded anechoic chamber.
Furthermore, all data collection procedures during outdoor measurements strictly adhered to ethical guidelines. 
Passive network monitoring and data collection were carried out solely using our own devices, each equipped with a valid SIM card. 
The process of connecting COTS UEs to live operator networks and logging RRC and PHY-layer messages reflects standard UE behavior and does not disrupt normal network operations.
This study complies fully with the terms of service of the participating wireless carriers and does not raise any ethical concerns.

\section{Conclusions \& Future Work}
\label{sec:conclusions}

In this work, we present \system, the first comprehensive O-RAN compliant system for detecting RBS that attempt downgrade attacks, deployed directly within the Near-RT RIC.
\system\ integrates an extended KPM Service Model and a custom xApp featuring ML-based anomaly detection. 
It enables real-time identification of RBS threats within a given area by classifying telemetry data based solely on UE ML1 logs and Measurement Reports.
The proposed extension to the 3GPP KPM Service Model allows for richer UE and E2 node-derived telemetry, enhancing detection capabilities. 
To validate \system\ under both controlled and real-world conditions, we built a dedicated testbed to verify the threat model and supplemented it with real-world measurements across commercial networks.
Among the models evaluated, the Variational Autoencoder achieved the best performance, with 99.5\% Accuracy and a False Positive Rate of just 0.6\%. 
Additionally, our system demonstrates low CPU and memory overhead, making it practical for deployment in production O-RAN environments.
As future work, we aim to extend the system to cover a wider range of attacks and further enhance the KPM Service Model to support richer telemetry. 
Beyond anomaly detection, we plan to explore ML-based resource optimization within the RAN and enable collaborative decision-making through integration with rApps at the SMO level.

\bibliographystyle{unsrt}
\bibliography{IEEEabrv,conference_101719}

\end{document}